# Mechanism of Structural Colors in Binary Mixtures of Nanoparticle-based Supraballs[†]

Christian M. Heil,[1,‡] Anvay Patil,[2,ϒ,‡] Bram Vanthournout,[3] Saranshu Singla,[2] Markus Bleuel,[4,5] Jing-Jin Song,[6] Ziying Hu,[7] Nathan C. Gianneschi,[7,8] Matthew D. Shawkey,[3] Sunil K. Sinha,[9] Arthi Jayaraman,[1,10,*] and Ali Dhinojwala[2,*]

[1]*Department of Chemical and Biomolecular Engineering, University of Delaware, 150 Academy St, Newark, Delaware 19716, USA.*

[2]*School of Polymer Science and Polymer Engineering, The University of Akron, 170 University Ave, Akron, Ohio 44325, USA.*

[3]*Evolution and Optics of Nanostructures Group, Department of Biology, Ghent University, Ledeganckstraat 35, Ghent 9000, Belgium.*

[4]*NIST Center for Neutron Research, National Institute of Standards and Technology, Gaithersburg, Maryland 20878, USA.*

[5]*Department of Materials Science and Engineering, University of Maryland, 4418 Stadium Dr, College Park, Maryland 20742, USA.*

[6]*Department of Materials Science & Engineering, University of California San Diego, 9500 Gilman Dr, La Jolla, California 92093, USA.*

[7]*Department of Chemistry, Northwestern University, Evanston, Illinois 60208, USA.*

[8]*Department of Materials Science and Engineering, Department of Biomedical Engineering, Department of Pharmacology, International Institute of Nanotechnology, Simpson-Querrey Institute, Chemistry of Life Processes Institute, Lurie Cancer Center, Northwestern University, Evanston, Illinois 60208, USA.*

[9]*Department of Physics, University of California San Diego, 9500 Gilman Dr, La Jolla, California 92093, USA.*

[10]*Department of Materials Science and Engineering, University of Delaware, 201 DuPont Hall, Newark, Delaware 19716, USA.*

[ϒ]Present address: *CertainTeed LLC, 20 Moores Road, Malvern, Pennsylvania 19355, USA.*

[‡]C.M.H. and A.P. contributed equally to this work.

[*]Corresponding authors: ali4@uakron.edu (A.D.), arthij@udel.edu (A.J.)

[†]Electronic supplementary information (ESI) available upon request.

**Abstract**

Inspired by structural colors in avian species, various synthetic strategies have been developed to produce non-iridescent, saturated colors using nanoparticle assemblies. Mixtures of nanoparticles varying in particle chemistry (or complex refractive indices) and particle size have additional emergent properties that impact the color produced. For such complex multi-component systems, an understanding of assembled structure along with a robust optical modeling tool can empower scientists to perform intensive structure-color relationship studies and fabricate designer materials with tailored color. Here, we demonstrate how we can reconstruct the assembled structure from small-angle scattering measurements using the computational reverse-engineering analysis for scattering experiments (CREASE) method and then use the reconstructed structure in finite-difference time-domain (FDTD) calculations to predict color. We successfully, *quantitatively* predict experimentally observed color in mixtures containing strongly absorbing melanin nanoparticles and demonstrate the influence of a single layer of segregated nanoparticles on color produced. The versatile computational approach presented in this work is useful for engineering synthetic materials with desired colors without laborious trial and error experiments.

**Table of contents graphic:**

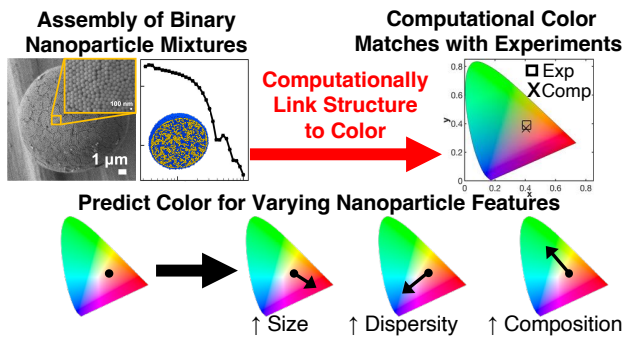

**Synopsis:**
The presented CREASE-FDTD approach provides quantitative color prediction for complex binary nanoparticle-based materials, enabling high-throughput material testing to design programmable colors.

**Introduction:**

The diverse array of coloration in nature (especially avian species)[1] has inspired advances in synthetic color fabrication.[2, 3] These structural colors arise from the spatial organization of the nanostructured material and, unlike pigment-based colors, are more resistant to color degradation. Materials with consistent, periodic nanostructures tend to form iridescent colors where the color varies based on the viewing angle, while those with only short-range ordering (long-range disordered) typically produce non-iridescent colors.[3, 4] To mimic these natural non-iridescent structural colors, researchers have utilized self-assembly of polymeric nanoparticles (such as polystyrene[5-7] and synthetic melanin[8, 9]) or inorganic nanoparticles (like silica[10-12]) to form amorphous assemblies. For the one-component nanoparticle assemblies studied, coloration can be varied by controlling the nanoparticle structure (size, dispersity, and packing) and optical properties (complex refractive index).[7, 13-15] Furthermore, binary nanoparticle mixtures provide increased structural diversity and optical responses. For instance, binary nanoparticle assemblies have been employed to produce isotropic, angle-insensitive structural colors by breaking long-range spatial periodicity.[5, 16-19] Research has shown that adjusting the nanoparticle size ratio between the components[20, 21] and the composition[18, 22] generates a myriad of diverse structures and structural colors.

While most optical research on disordered nanoparticle assemblies has focused on non-absorbing materials, many avian species generate structural colors by the arrangement of highly absorbing melanosomes (melanin-containing organelles) in a non-absorbing keratin matrix.[23, 24] Melanin is a particularly interesting bio-inspired pigment material for optical applications due to its *high refractive index* and *broadband absorption*.[25, 26] By varying the arrangement of the melanosomes from weakly segregated (*e.g.*, ravens) to strongly segregated (*e.g.*, California quail), these animals produce a wide variety of colors.[27] Inspired by melanin's widespread use in natural systems, researchers have developed synthetic melanin mimics for a wide range of applications, particularly in the optics field.[2, 28, 29] In addition to the bulk structure influencing the coloration, research on synthetic systems has demonstrated that binary nanoparticle assembly in the presence of an interface can result in differing bulk and surface composition and segregation.[30, 31] A recent study has shown that mixing melanin and silica nanoparticles during a reverse-emulsion assembly process leads to the formation of a melanin shell in the resulting supra-structure (termed supraball).[31] The shell formation on the supraball reflectance was further qualified to demonstrate that the surface properties strongly impact the reflectance properties.[15] Thus, binary nanoparticle mixtures provide easily adjustable methods to expand the accessible color space by controlling the composition, the extent of interparticle mixing, and surface segregation. With a plethora of parameters known to influence color generation, there is a need for a method to predict/model color generation to aid in the design of nanoparticle assemblies with specific coloration.

Typically, optical modeling of nanoparticle assemblies has been performed using semi-empirical methods such as diffusion theory,[32-34] single-scattering approximations using Mie theory,[6, 7, 13] and Monte Carlo-based multiple scattering models.[35, 36] A drawback of some semi-empirical models is they require an iterative process whereby one adjusts system properties (*e.g.*, nanoparticle size, density, and refractive index), evaluates the alteration's impact on optical properties, and further modifies to achieve suitable optical modeling prediction match to the experimental results. Furthermore, those optical modeling techniques treat the nanoparticle assemblies with effective medium approximations, rendering the approaches unsuitable for complex systems with heterogeneity especially for multi-component nanoparticle assemblies. To the best of our knowledge, previous approaches employing optical modeling to predict structural color have focused on one-component nanoparticle assemblies. To address the shortcomings in other optical modeling approaches, we utilize the finite-difference time-domain (FDTD) method because FDTD is a first-principle technique, suitable for performing *quantitative* optical modeling of complicated nanoparticle assemblies.[37, 38] Previous work has successfully utilized an FDTD-based approach to predict structural coloration of binary chemistry nanoparticle assemblies with large refractive index contrasts between the different nanoparticle chemistries.[15] Using FDTD, one can calculate the absorbance, transmittance, and reflectance of materials by discretizing the material onto a spatiotemporal grid and numerically solving Maxwell's curl equations for each grid point at each time step. As such, to predict the color production from binary assemblies using FDTD, one must understand the bulk and surface structural arrangement and possess a method to reconstruct a structure representative of the underlying nanoparticle assembly.

Small-angle scattering (SAS) and electron microscopy are common approaches to obtain structural information.[39-47] However, electron microscopy typically is only performed for a small subset of supra-particles and can only probe limited length-scales compared to SAS.[41, 42, 48] SAS techniques, small-angle neutron and X-ray scattering (SANS/SAXS), are well suited to obtain ensemble-averaged bulk structural information across a wide length-scale.[48] The output of the SAS experiment is a scattering intensity profile, $I(q)$, as a function of the wavevector, $q$, with the scattering intensity in the inverse-distance space.[40] Thus, interpretation of the scattering intensity is necessary and commonly performed using analytical models.[48, 49] However, those models do not provide a 3D structural reconstruction that would enable optical modeling using a first-principle technique, such as the FDTD method. Instead, we utilize the recently developed computational reverse-engineering analysis for scattering experiments (CREASE) method that has been extensively validated for nanoparticle assemblies.[50, 51] Previous work has demonstrated how combining the CREASE and FDTD methods allows a user to accurately predict the structural color of one-component nanoparticle assemblies.[51]

In this work, we apply this combined CREASE-FDTD approach to binary nanoparticle assemblies forming supraballs because using mixtures with varying particle sizes and refractive indices produce a wider spectrum of colors. However, mixtures of nanoparticles can also lead to complexity related to bulk and surface phase segregation, necessitating non-trivial structural characterization. We first demonstrate that CREASE determines 3D structures with computed scattering profiles similar to those obtained using contrast matching SANS experiments. The 3D structure determined from CREASE is then used in the FDTD method to predict reflectance profiles and colors. The close agreement between computational and experimental reflectance profiles for a binary mixture of silica and melanin nanoparticles at different compositions strengthens the applicability of this combined CREASE-FDTD approach to relate the internal structure of complex multicomponent mixtures of particles of different shapes, sizes, and chemistries to their optical responses. We then further investigate and quantify how segregation of a surface layer of melanin particles impacts the structural color produced. We conclude by demonstrating how the CREASE-FDTD approach can also be used to explore novel, yet-to-be synthesized systems by predicting the color response for various perturbations introduced to the material design. This work validates the combined CREASE-FDTD approach as being well-suited to characterizing optical properties of disordered nanoparticle assemblies and demonstrates how the approach can be utilized to design colors on demand for applications in paints, cosmetics, and food coloring.

**Results and Discussion**

*Structural Color Diversity Produced by Binary Nanoparticle Mixtures*

      Differences in composition and phase morphology of binary mixtures of absorbing and non-absorbing species of similar size provide an easy method to control the production of structural colors. **Figure 1** demonstrates the relative sensitivity of structural color to these two design parameters. We perform color reflectance calculations using FDTD on *in-silico* binary nanoparticle mixtures of absorbing melanin and non-absorbing silica nanoparticle-based supra-assemblies (called supraballs) generated using coarse-grained molecular dynamics simulations under spherical confinement (**Figure 1A**). The rich diversity of colors produced (**Figure 1B**) demonstrates the necessity of a thorough knowledge of not only the relative composition of the two nanoparticle types in a binary mixture, but also the phase segregation/extent of mixing. As the extent of mixing decreases, the nanoparticle assemblies have increasingly segregated domains, demonstrating the need for an experimental and computational approach that can provide information about the supraballs' internal morphologies needed for predicting color.

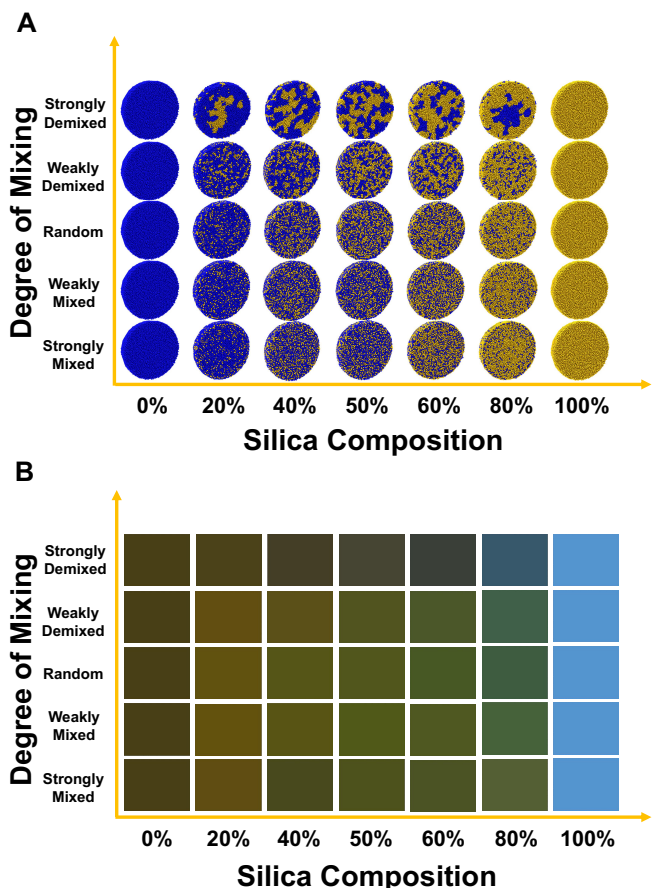

*Figure 1*. Effect of monodisperse binary nanoparticle mixture (220 nm diameter melanin and 220 nm diameter silica; melanin: blue spheres; silica: yellow spheres) composition and mixing state on the supraball color reflectance. (A) Visualizations of the cross-section of binary mixture supraballs with varying levels of particle mixing in the increasing order from top to bottom and *varying relative proportion of silica in the increasing order from left to right.* (**B**) *Corresponding structural colors, represented as RGB color panels, of the binary mixture supraballs.*

*CREASE Reconstruction from SANS Data*

The SANS experiments were conducted for three different mixtures (1:4, 1:1, or 4:1 melanin:silica by volume) of nanoparticles to produce suspensions of 5-10 μm-size supraballs (**Figure 2**). To obtain information on morphology within the supraballs, we needed to isolate the scattering contribution of individual components within the supraball. The contrast matching technique in SANS (for details see ESI **Section 1**) facilitates the collection of the total scattering from both melanin and silica nanoparticles (denoted as the non-contrast-matched (NCM) condition) and the scattering from silica nanoparticles only (denoted as melanin contrast-matched (MCM) condition). The MCM condition is achieved by adjusting the deuterated and hydrogenous levels of the solvent medium, so the melanin particle scattering is matched to the background scattering resulting in the coherent scattering solely from silica nanoparticles. Off-the-shelf analytical scattering models are not readily available to model such densely packed, binary nanoparticle mixture system. Therefore, we require an alternative generic approach like CREASE that analyzes scattering results from such multicomponent structures and produces a 3D structural reconstruction to perform optical simulations.

We apply the recently developed CREASE method[50] to reconstruct the binary supraball structures in **Figure 2**. In **Figure 2A**, we show that the CREASE method takes both the experimental NCM and MCM scattering intensity profiles as input and outputs a 3D structure, whose computed scattering profile closely matches the experimental input. The CREASE method[50] represents the nanoparticles' structure as a set of genes describing the nanoparticles' mean diameter and dispersity, nanoparticle concentration, spatial arrangement of the nanoparticles, and the size of the reconstructed structure (or reconstruction size). CREASE employs a genetic algorithm to optimize the genes to ones that correspond to structure(s) with a computed scattering intensity, $I_{comp}(q)$, that most closely matches the target *i.e.*, experimental scattering input, $I_{expt}(q)$. As we showed in our previous study,[51] we can leverage the gene-based nature of CREASE to quickly determine the values of the genes using a smaller reconstruction size (~6 μm in diameter) before using those genes to reconstruct the larger 3D structure with similar dimensions to the experimental supraballs (~10 μm in diameter). This two-step approach reduces the computational time required while obtaining output structures with strong agreement between the $I_{expt}(q)$ and $I_{comp}(q)$ for both NCM (*top*) and MCM (*bottom*) conditions for all binary mixture supraballs (**Figure 2B-2D**).

As previously mentioned, SANS provides information on the bulk structural arrangement, but the binary mixture supraballs in this work have a surface segregated layer of melanin (**Figure 2E**). Previous work demonstrated that melanin forms a surface layer on the supraballs as a direct result of the supraball processing.[31] During the emulsion assembly, the melanin nanoparticles have a greater contact angle than the silica nanoparticles at the water-octanol interface. The larger contact angle and the shrinking spherical geometry during the emulsion assembly cause the formation of the melanin layer. As was previously shown,[15] due to distinct optical properties of melanin and silica, this melanin shell layer has a strong influence on the overall supraball reflectance. Thus, to correctly represent the experimental supraball systems with the surface layer of melanin, we added a layer of melanin nanoparticles onto the top and bottom of the CREASE reconstruction of the bulk structure (**Figure 2E**). Computational work demonstrated that as the supraball diameter increases, the bulk composition within the supraball approaches the overall composition, and for the supraballs of interest in this work (~10 μm in diameter), the bulk composition and overall composition are nearly identical.[30] Thus, the addition of the melanin layer to the top and bottom of the CREASE reconstructed structure does not significantly impact the overall composition and matches the predicted composition observations.

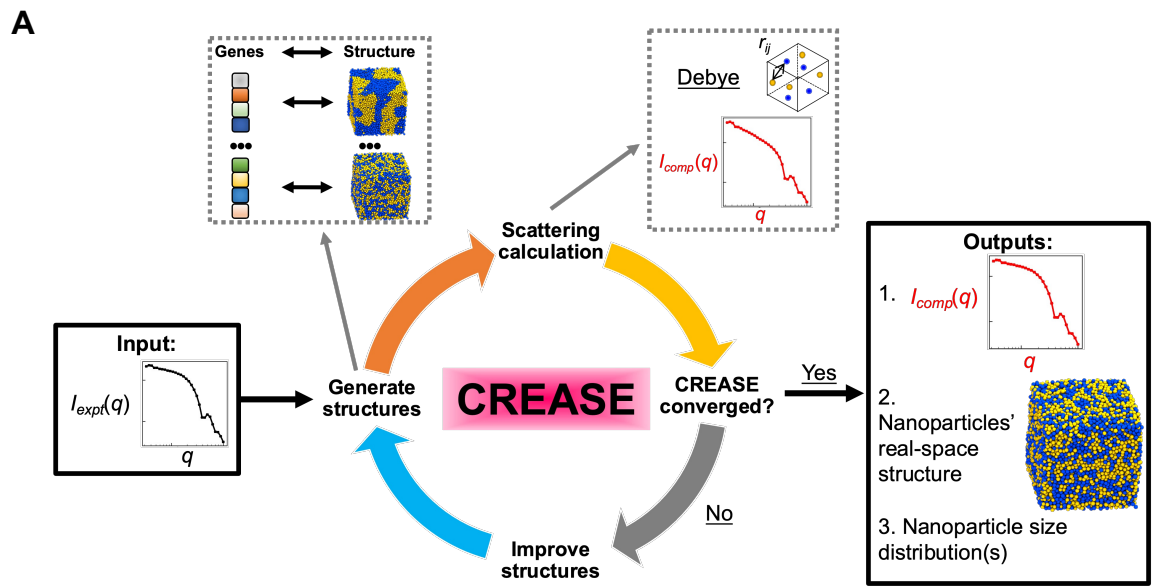

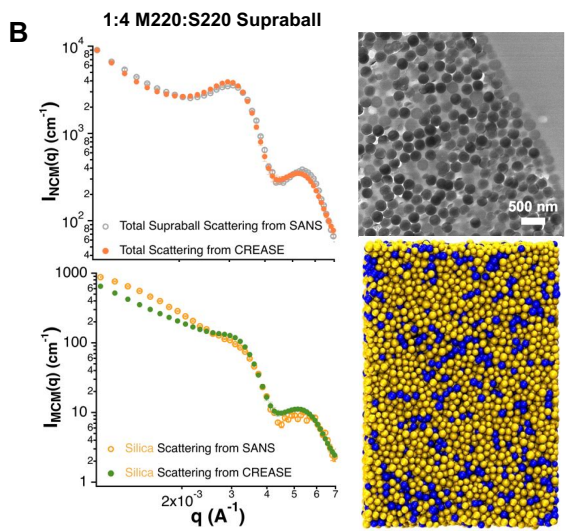
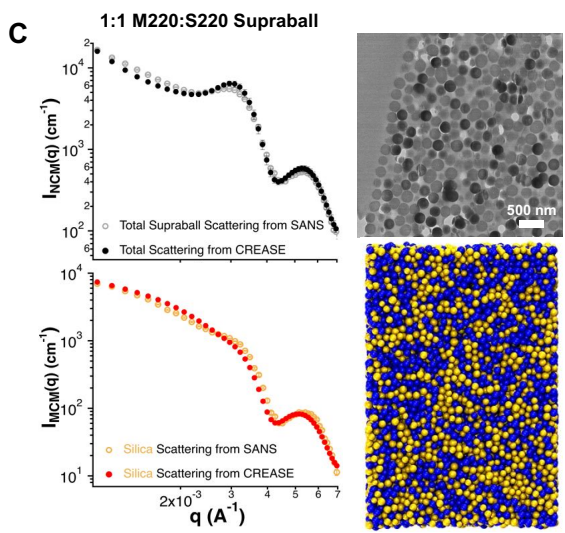
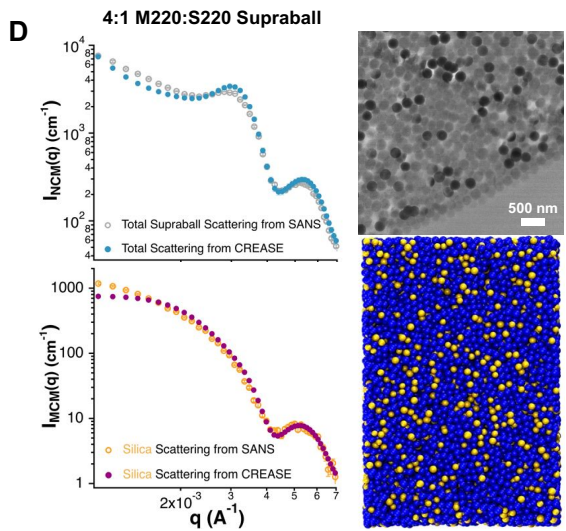
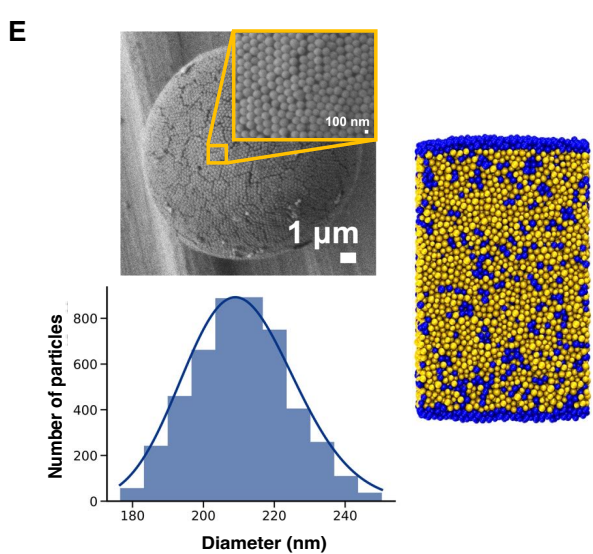

*Figure 2*. Applying the CREASE method to reconstruct the binary nanoparticle mixture assembly structure from SANS profiles. (*A*) Schematic describing the CREASE method operation. (*B-D*) SANS plot of I as a function of q for non-contrast-matched (NCM) (left top; gray dots) and melanin contrast-matched (MCM) (left bottom; yellow dots) condition of binary mixture supraballs overlaid with the CREASE output structures' scattering profile for NCM and MCM condition for *B*) 1:4, *C*) 1:1, and *D*) 4:1 melanin:silica compositions. *B*) plots the NCM and MCM from CREASE as orange and green, *C*) colors the NCM and MCM from CREASE as black and red, and *D*) shows the NCM and MCM from CREASE as blue and purple, respectively. *B-D*) show, on the right, transmission electron micrograph of the cross-section of a representative binary mixture supraball (right top; melanin: lighter spheres, silica: darker spheres; scale bar: 500 nm) and VMD visualization of the central portion (3 μm × 3 μm × 10 μm) of reconstructed 3D binary nanoparticle mixture assembly with yellow spheres representing silica chemistry and blue spheres representing melanin (right bottom). (*C*) Scattering profile and CREASE results originally from Ref.[50] (*E*) Scanning electron micrographs of a representative 1:4 binary mixture supraball and supraball surface (left top; scale bars are 1 μm and 100 nm, respectively) and lognormal size distribution of the melanin nanoparticles used to form the surface segregated shell after CREASE reconstruction (left bottom). VMD visualization of the of the central portion (3 μm × 3 μm × 10 μm) of reconstructed 3D binary nanoparticle mixture assembly with the added melanin shell layer (right).

*Optical Modeling of Binary Supraballs using FDTD Method*

The CREASE's 3D structural reconstruction provides the x, y, and z coordinates of all the nanoparticles within the supraball and this information can then be used as an input to calculate the scattering of light using FDTD methods. In **Figure 3**, we demonstrate that optical modeling using the FDTD method on CREASE's output 3D structures provides computed reflectance spectra that closely matches the experimental reflectance spectra. For the 1:4 melanin and silica binary nanoparticle mixture supraballs in **Figure 3A**, the computed and experimental reflectance spectra are indistinguishable within error. The visual perception of the computed reflectance, shown by the RGB color panel, closely matches with the experimental RGB color. The chromaticity coordinates of both experimental and computed colors are marked on the CIE 1931 color chart to indicate their relative closeness. Color difference ($\Delta E$) analysis reveals that the computed and experimental colors are similar, only ~0.9 times the average just noticeable difference (JND) value, indicating that fewer than 50% of observers can distinguish between them. **Figure 3B** provides the comparison for the 1:1 binary mixture supraballs. The equal composition system likewise achieves a close quantitative agreement between the computed and experimental reflectance with a $\Delta E$ of only ~1.9 times the average JND value. The final binary mixture supraball system with 4:1 composition, majority component melanin, is shown in **Figure 3C**, and the combined CREASE-FDTD approach achieves a computed reflectance spectra with color difference of ~1.4 times the average JND value. Thus, for these complex binary nanoparticle mixture systems, we confirm the CREASE output achieves a close structural match to the experimental systems using optical modeling comparison. **Figure 3** demonstrates our capability to consistently produce reconstructed structures that mimic the relevant color-producing structural intricacies in experimental systems, enabling researchers to develop structure-color relationships as we demonstrate in the proceeding section.

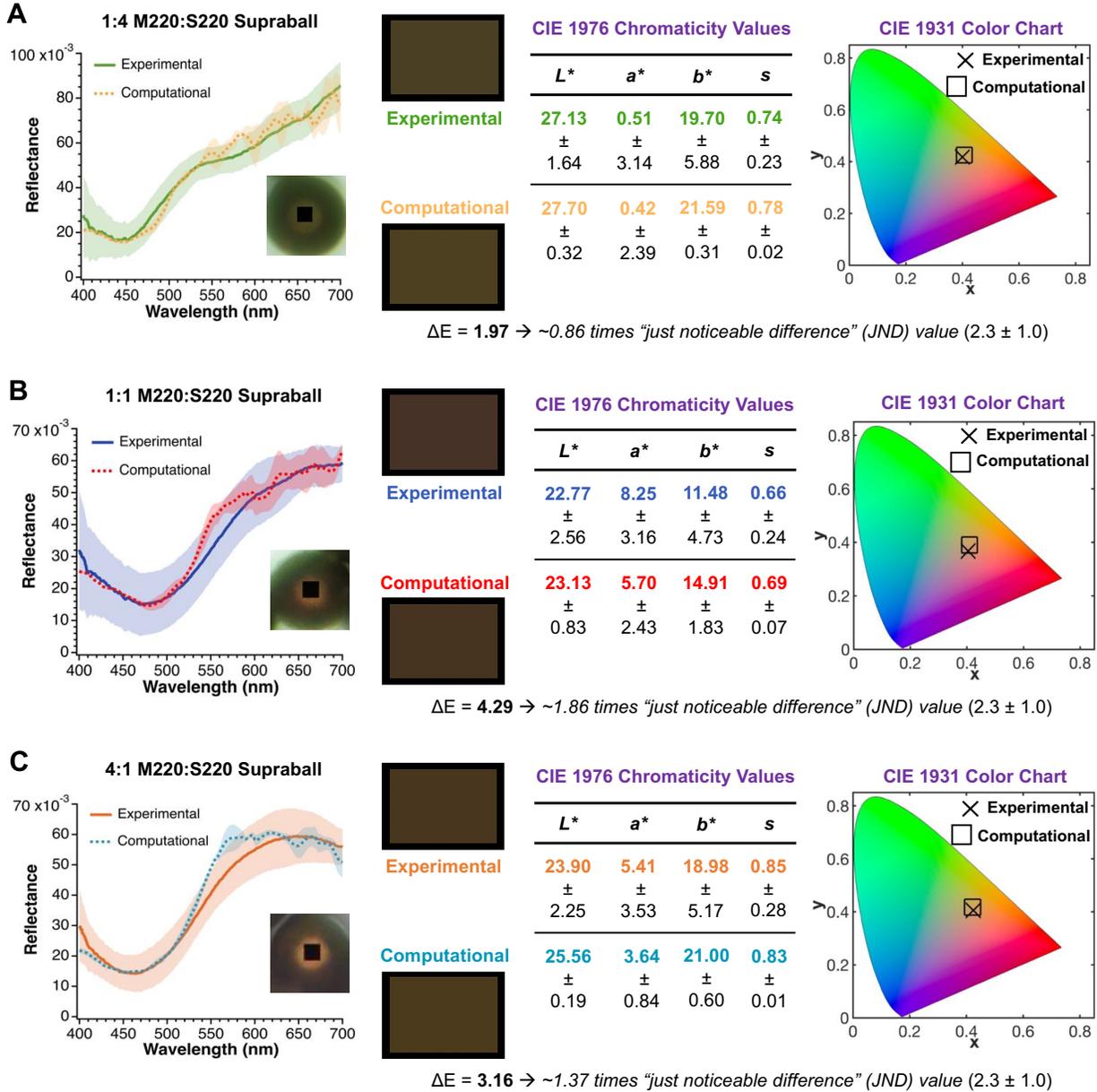

*Figure 3*. Optical modeling and color analysis comparison between experimental supraballs and FDTD calculations on the CREASE output structures. (*A*) Reflectance spectra (solid curve: experimental; dashed curve: computed), RGB color panel, CIE 1976 chromaticity values, and CIE 1931 chromaticity co-ordinates' comparisons for the 1:4 melanin:silica binary mixture supraball system. The quantitative difference between FDTD and experimental colors is given by a color difference (*ΔE*) value that is ~0.9 times the average just noticeable difference (JND) value. The black box in the inset of the optical micrograph of corresponding supraball represents the size of the area (3 μm × 3 μm) probed during optical measurements using microspectrophotometer. (*B*) Similar to (*A*) but for 1:1 melanin:silica binary mixture supraball system with a *ΔE* value that is ~1.9 times the average JND value. (*C*) Similar to (*A*) but for 4:1 melanin:silica binary mixture supraball system with a *ΔE* value that is ~1.4 times the average JND value.

*Structure-Color Property Elucidation Using Combined CREASE-FDTD Approach*

The excellent agreement of the combined CREASE-FDTD approach to model optical properties of a complex binary mixture of nanoparticles, we now demonstrate how this approach can help us understand the influence of size dispersity and surface segregation on purity of structural colors. First, we investigate the effect of the melanin shell layer on the binary mixture supraball reflectance (**Figure 4**). Experimentally, this would be similar to engineering the emulsion interfacial properties by either changing the solvents used during the emulsion assembly or modifying the melanin/silica particle surface to avoid the melanin surface segregation. For the 1:4 system in **Figure 4A**, we find a significant change in the reflectance profile when the melanin shell is included, resulting in suppression of the prominent reflectance peak around ~560 nm, drop in the lightness factor by ~24%, a ~51% increase in the yellow component, and a ~108% increase in the red component (negative a* to positive a* transition) of the color produced. The introduction of the melanin surface layer results in a $\Delta E$ of ~5.5 times the average JND value with nearly double the color saturation of the supraballs without the melanin layer. Interestingly, **Figure 4B**, showing the results for the 1:1 composition, illustrates that the addition of the melanin shell does not consistently result in a redder color (insignificant change in the red component). The reconstructed structures with melanin shell exhibit a ~23.0% drop in lightness factor compared to structures without melanin shell, with a negligible change in the color saturation. The distance traversed by the chromaticity co-ordinates on the CIE 1931 color chart, upon addition of the melanin shell layer, is marginal; the $\Delta E$ value of the two colors is ~3.7 times the average JND value. This finding contrasts with previous work[15] that examined *monodisperse* binary chemistry melanin and silica supraballs with a melanin shell and without one and found the melanin shell caused a substantially redder color. The incorporation of dispersity in particle sizes is the likely cause for the differing trends seen with the addition of a melanin shell layer. Finally, we consider the 4:1 system that we expect to have negligible differences in the color and reflectance with and without the melanin shell layer as the composition is majority melanin particles already (**Figure 4C**). The reflectance curves from the nanoparticle assembly with and without the melanin shell are the most similar, compared to the other compositions considered, with only slight differences. However, the CIE 1931 color chart indicates that the resulting colors are quite different with the addition of the melanin shell, increasing the yellow component by 38% and decreasing the red component by ~63%, in contrast to the 1:4 system in **Figure 4A**. Quantitatively, the two colors have a $\Delta E$ of ~3.7 times the average JND value with merely 15% improvement in color saturation for systems with melanin shell. **Figure 4** clearly demonstrates addition or removal of the melanin shell can provide an additional way to fine-tune the structural color produced; however, the trends may not be universally consistent suggesting a high-throughput approach, such as using this CREASE-FDTD method, is warranted.

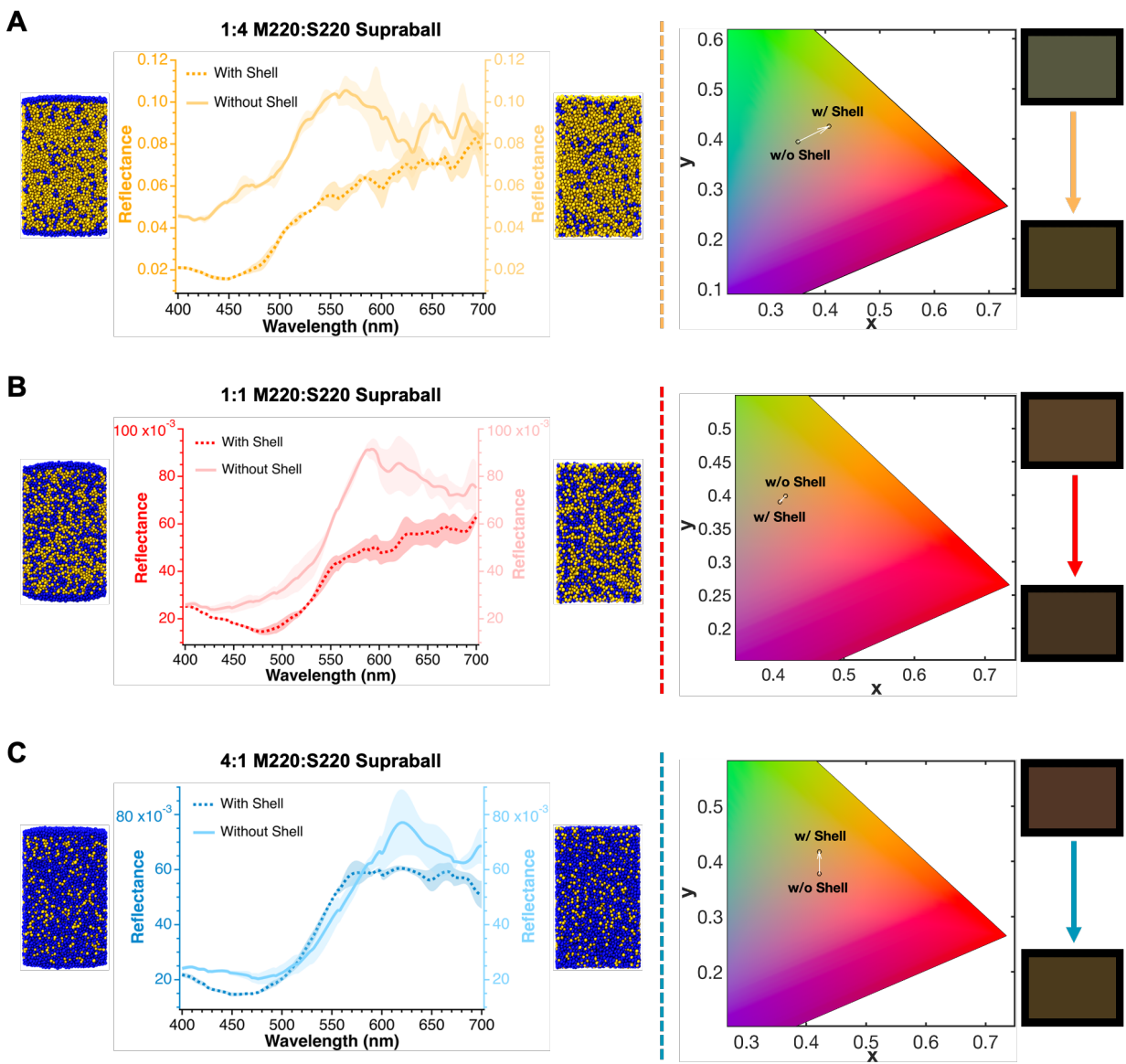

*Figure 4*. Optical modeling and color analysis comparison between FDTD calculations on the CREASE output structures with and without the melanin shell. *(A-C)* Reflectance spectra (solid curve: with melanin shell; dashed curve: without melanin shell), structure visualizations, CIE 1931 chromaticity co-ordinates', and RGB color panel comparisons for the *A)* 1:4, *B)* 1:1, and *C)* 4:1 melanin:silica binary mixture supraball systems.

We further investigate the effect of the melanin shell layer on the 1:4 binary mixture supraball reflectance by varying the average nanoparticle size and dispersity (**Figure 5**). Experimentally, researchers have found approaches to stratify nanoparticle assemblies producing both larger nanoparticles-on-top and smaller nanoparticles-on-top.[52] Furthermore, one could consider a double-emulsion or two-step emulsion approach to tune the melanin shell nanoparticle characteristics separately from the bulk size distribution.[53, 54] **Figure 5A** depicts the effect of changing the lognormal average diameter from the mean, experimental value (~210 nm; **Figure 2**) while maintaining a consistent dispersity (~7%; **Figure 2**) on the reflectance and corresponding color. The reflectance spectra achieve *qualitatively* similar shapes with a linear shift in the trough (at low wavelengths) based on the increase/decrease in the average shell particle diameter. While the color chart and RGB panels illustrate that the net change is toward a redder (greener) color when increasing (decreasing) the average particle size, we note that the overall path from mean minus 20 nm to mean plus 20 nm is surprisingly not linear. We next consider how changing the lognormal dispersity with a constant average diameter (~210 nm) affects the resulting structural coloration (**Figure 5B**). We select three dispersity values of 1%, 10%, and 20% to illustrate the overall trend that increasing the shell particle size dispersity results in a flatter reflectance curve and a grayer (more neutral) color. The change in color chart indicates that the change in dispersity up to 20% linearly affects the blue-shift of the color in accordance with previous work[15] that investigated changing the bulk dispersity in one-component melanin supraballs. However, this current work demonstrates that the same net color effect (blue-shift) is achievable by *solely* adjusting the melanin shell layer, not the whole supraball.

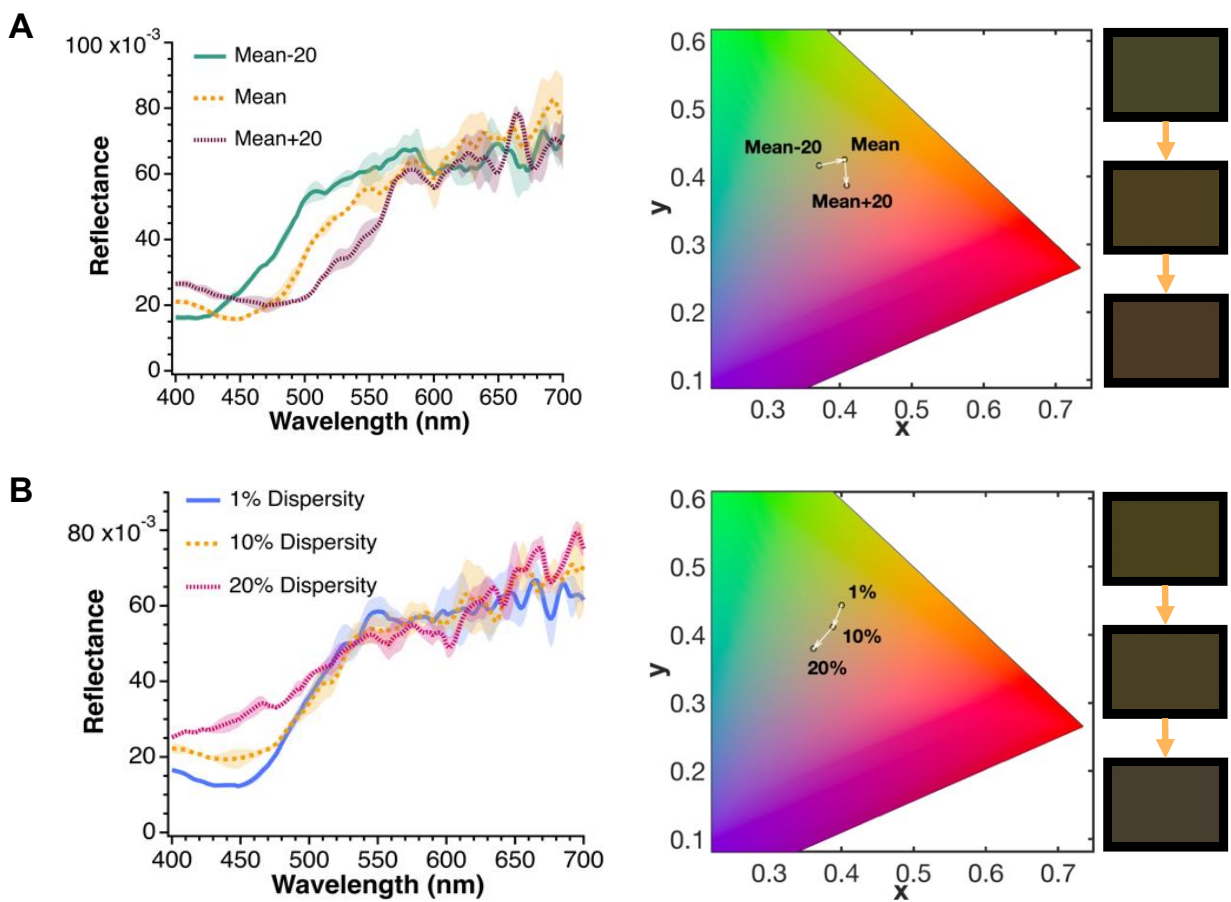

*Figure 5*. Optical modeling and color analysis comparison between FDTD calculations on the 1:4 composition CREASE output structures as the nanoparticle size and dispersity are varied in the melanin shell layer. (**A**) Reflectance spectra, as the lognormal mean diameter of the melanin shell nanoparticles, is increased (purple, dashed line) or decreased (green, solid line) from the mean used in Figure 2-4 (yellow, dotted line). The CIE 1931 color chart and RGB color panels provide the predicted color change when changing the average diameter of the melanin shell particles. (**B**) Reflectance spectra, as the lognormal of the melanin shell nanoparticles, is varied from 1% (blue,

*solid line) to 10% (yellow, dotted line) to 20% (pink, dashed line). For reference, the experimental 1:4 system has an average size dispersity of ~7%. The CIE 1931 color chart and RGB color panels provide the predicted color change when changing the size dispersity of the melanin shell particles.*

**Conclusions**

This work shows how a combination of CREASE and FDTD provides a platform to model structural colors for mixtures of nanoparticle-based supra-assemblies as a function of size, dispersity, phase morphology, and strong absorbing optical properties. We demonstrate that this combined method produces reconstructed three-dimensional structures of nanoparticle assemblies whose scattering profiles match those obtained using SANS measurements and possess structural colors that match the colors measured using optical spectroscopy. The combined CREASE-FDTD approach demonstrates and quantifies the influence of the surface segregation of melanin nanoparticles occurring in the experimental supraball assemblies on the resulting color. We also show how changing the melanin surface layer characteristics (diameter or dispersity), beyond what is produced in current experiments, leads to significant shifts in the resulting structural color. Moving forward, a combination of CREASE-FDTD and multi-scale modeling can further extend this method to study optical responses of much larger assemblies of supraballs.[55] The CREASE-FDTD approach can also serve as a high-throughput method to design programmable colors for applications in the areas of coatings, paints, and cosmetics. Ultimately, the application of CREASE and FDTD will allow researchers to better design complex materials for applications in other regions of the electromagnetic spectrum.

**Experimental Methods**

*Nanoparticle Synthesis*

Silica nanoparticles were synthesized using a modified Stöber process.[56] The synthetic melanin nanoparticles were prepared using a previously reported,[31] base-catalyzed, auto-oxidative polymerization of dopamine monomer (Sigma Aldrich), a commonly accepted synthetic moiety of natural eumelanin (the most abundant form of melanin). The synthesis step involved constant stirring of a mixture of dopamine, water, ethanol, and ammonia solution ($NH_4OH$; Sigma Aldrich - 28 to 30 wt.%) under ambient conditions. To inspect particle morphology, we drop-casted dilute silica and melanin nanoparticle suspensions onto carbon-coated copper grids (FCF300-Cu; Electron Microscopy Sciences) for transmission electron microscopy (JEM-1230, JEOL Ltd.).

*Supraball Fabrication and Characterization*

The binary mixture supraballs were fabricated following our previous reverse-emulsion assembly process.[31, 57] Typically, the preparation of the reverse-emulsion involved adding 30 µl of aqueous solution of silica and melanin nanoparticles (30 mg/mL each; 1:4 melanin:silica mixture composition: 20.8 µl of silica + 9.2 µl of melanin nanoparticle suspensions, 1:1 melanin:silica mixture composition: 10.8 µl of silica + 19.2 µl of melanin nanoparticle suspensions, 4:1 melanin:silica mixture composition: 3.7 µl of silica + 26.3 µl of melanin nanoparticle suspensions ) to 1 mL of anhydrous 1-octanol (Sigma-Aldrich) and vortex mixing the mixture at a speed of 1600 rpm for 2 min followed by 1000 rpm for 3 min. The newly formed supraballs were precipitated, the supernatant was extracted, and the samples were dried at 60°C.

The binary mixture supraballs were visualized using a field-emission scanning electron microscope (JSM-7401F, JEOL Ltd.). To inspect the internal structure of the supraball, the dried supraball powder was embedded in an epoxy resin (EMbed 812) in a block mold and cured at 60°C for *ca.* 16 h. The cured block was trimmed using a Leica S6 EM-Trim 2 (Leica Microsystems) to give a sharp trapezoidal tip. Later, 80 nm-thick slices were cut from the trimmed block using a diamond knife (Diatome Ltd.) on a Leica UC7 ultramicrotome. These slices were loaded onto carbon-coated copper grids (FCF300-Cu; Electron Microscopy Sciences) for transmission electron microscopy.

The dried binary mixture supraballs were placed on a Piranha-cleaned silicon wafer (Silicon Inc.) and were evaluated for their color reflectance property at normal incidence using a CRAIC AX10 microspectrophotometer (CRAIC Technologies, Inc.). A 50x objective and a 75-W Xenon short arc lamp (Ushio UXL-75XE) for the white light source were employed. Silver mirror was used as a reflectance standard (white standard) for instrument calibration. The lamp-off condition was used to record dark counts for black background subtraction. The experimental reflectance spectra reported were an average of 15 measurements with the error bars represented by the standard deviation. To account for substrate effect in the reflectance measurement (unlike in the computational case where the supraball was in an air background, $n=1$), the experimental reflectance spectra were subtracted by a constant factor (1:4 mixture composition: +0.035, 1:1 mixture composition: +0.025, 4:1 mixture composition: +0.028) and plotted to the same scale as the computed reflectance using a multiplier factor (1:4 mixture composition: 1.125, 1:1 mixture composition: 0.75, 4:1 mixture composition: 0.72). We have previously demonstrated[51] that it is essential to account for the substrate effect in our analyses. All the color analyses for experimental systems were performed on the corrected reflectance datasets.

*SANS Measurements*

All the SANS experiments were conducted at the National Institute of Standards and Technology Center for Neutron Research (NIST CNR or NCNR). To obtain the primary nanoparticle shape, size, and polydispersity (form factor description), we performed SANS on dilute aqueous suspensions of melanin and silica nanoparticles (prepared in deuterated water to avoid incoherent scattering). The standard configurations of the beamlines were used to run the form factor measurements *i.e.*, *a*) for melanin, at the NGB 30m SANS instrument, 1, 4, and 13 m sample-to-detector distances were used with 6 Å neutrons while the Lens setup for low-*q* used 8 Å at 13 m, and *b*) for silica, at the vSANS instrument, the high-*q* setup used 6 Å neutrons with front and middle detector carriages set at 1.1 m and 5.1 m, respectively, from the sample while the low-*q* setup used 11 Å neutrons with front and middle detector carriages set at 4.6 m and 18.6 m, respectively, from the sample.

The supraball scattering curves, both non-contrast-matched (NCM; total scattering contribution by both melanin and silica) and melanin contrast-matched (MCM; silica scattering only), were obtained by performing SANS measurements on supraball suspensions in 1-octanol with different levels of deuteration, at the vSANS beamline. The measurements started with supraballs suspended in 100% deuterated 1-octanol to acquire NCM scattering intensity profiles, followed by spiking the solvent medium with hydrogenous 1-octanol to reach the melanin contrast-matching condition *i.e.*, 60% deuterated and 40% hydrogenous (see ESI **Section 1**), to get MCM scattering intensity profiles. The instrumental configurations were the same as described previously for vSANS.[15]

The dilute primary nanoparticle and supraball suspensions were contained in quartz banjo cells (Product # 120-2mm and 120-1mm, respectively; Hellma USA) to avoid any undesired scattering contribution from the containers. The SANS experiments were conducted at ambient temperature and the measured intensities were corrected for background scattering and empty cell contributions. They were also normalized using a reference scattering intensity of a polymer sample of known cross-section. The reduction of raw SANS data was performed following a well-known protocol described by Kline.[58] The processed SANS form factor data were analyzed using SasView 5.0.4 (https://www.sasview.org). All the experimental scattering datasets and model/computed fits presented in this study were pinhole-smeared. No unexpected or unusually high safety hazards were encountered.

**Computational Methods**

*CG-MD Simulations*

Previous works have detailed the procedure to produce the simulated supraball structures.[30, 31] In this CG approach, the nanoparticles were represented as spherical particles (melanin: blue; silica: yellow) with the monodisperse diameter set at 220 nm. For the case of binary mixtures with varying relative proportions of melanin and silica in the mixture composition, all nanoparticles interacted with other nanoparticles through the colloid Lennard-Jones (cLJ) potential[59] with Hamaker constants set to achieve different degrees of particle mixing: strongly demixed supraballs ($A_{Mel-Mel}$ = 0.50 $k_BT$, $A_{Mel-Sil}$ = 0.08 $k_BT$, and $A_{Sil-Sil}$ = 0.50 $k_BT$), weakly demixed supraballs ($A_{Mel-Mel}$ = 0.25 $k_BT$, $A_{Mel-Sil}$ = 0.20 $k_BT$, and $A_{Sil-Sil}$ = 0.25 $k_BT$), randomly mixed supraballs ($A_{Mel-Mel}$ = 0.25 $k_BT$, $A_{Mel-Sil}$ = 0.25 $k_BT$, and $A_{Sil-Sil}$ = 0.25 $k_BT$), weakly mixed supraballs ($A_{Mel-Mel}$ = 0.20 $k_BT$, $A_{Mel-Sil}$ = 0.25 $k_BT$, and $A_{Sil-Sil}$ = 0.20 $k_BT$), and strongly mixed supraballs ($A_{Mel-Mel}$ = 0.08 $k_BT$, $A_{Mel-Sil}$ = 0.50 $k_BT$, and $A_{Sil-Sil}$ = 0.08 $k_BT$). The spherical supraball was formed by implementing a spherical wall to mimic the emulsion droplet interface with an attractive harmonic potential interaction between the spherical wall and the particles. The attractive harmonic potential was set using $\varepsilon$ = 500 $k_BT$ and using a cutoff set to the nanoparticles' radius. The characteristic simulation length $\sigma$ was 1.0 nm, the characteristic mass was the mass of the melanin nanoparticle (scaling silica nanoparticles' mass by its relative density), and the characteristic energy $\varepsilon$ was 1.0 $k_BT$.

To simulate the supraball formation, the particles were randomly inserted into a spherical region (radius ~13 μm) with an initial occupied volume fraction of 0.03. We utilized Langevin dynamics in the LAMMPS software package[60] to maintain system temperature and to mimic solvent effects. After the previously described equilibration,[30, 31] the spherical wall radius was decreased to model the shrinking emulsion droplet. We utilized a simulation timestep of 0.0025 τ (τ represents simulation time) until the close-packed supraball was formed with a final occupied volume fraction of ~0.6.

*CREASE Method*

The gene-based CREASE method determines 3D bulk structure from small-angle scattering experiments. The full details for the CREASE method can be found in previous work.[50]

For the binary mixtures in this study, CREASE takes as an input the experimental scattering profile of NCM samples $I_{expt-NCM}(q)$ and of MCM samples $I_{expt-MCM}(q)$ and utilizes a genetic algorithm-based approach to determine the nanoparticle structure. Here, the binary mixture composition was set to guide the reconstruction, however, the user does not need to input the exact values; they can provide an upper and lower bound of the composition to let CREASE converge the value. The nanoparticle size distribution was determined by CREASE and agreed with the results obtained from form factor measurements performed using SANS. CREASE initializes a population of individuals where each individual contains a set of genes that are related to the nanoparticle diameter and size dispersity, nanoparticle concentration, spatial arrangement of nanoparticles, and the number of nanoparticles in the 3D structure reconstruction. For each individual, CREASE calculates the computed scattering intensity $I_{comp-NCM}(q)$ and $I_{comp-MCM}(q)$ with the Debye scattering equation [61, 62]. CREASE calculates the fitness for each individual which quantifies the quality of the match between computed and experimental $I(q)$s. After each generation, each individual's set of genes are altered by performing mutations and combinations that either randomly change the gene or average two good gene values together, respectively. After the fitness converges, CREASE outputs the best 3D nanoparticle structure and its corresponding computed $I(q)$s (which closely matches the experimental $I(q)$s) for use in the FDTD calculations.

The CREASE method determines the bulk structure from small-angle scattering. However, the supraballs of interest are known to have a single segregated layer of melanin nanoparticles on the surface, and the optical properties of the supraballs are heavily influenced by the outermost layers of the supraball.[31] Thus, after we determine the bulk nanoparticle structure with CREASE, we must then add a layer of melanin particles to mimic the supraball surface before performing FDTD calculations. We start by creating a cubic arrangement of melanin 'shell' nanoparticles in a cubic arrangement above (+z direction) and below (-z direction) the bulk structure returned by CREASE. The shell nanoparticles have their sizes set using a lognormal distribution with the desired average nanoparticle diameter and dispersity. The lognormal distribution of nanoparticle diameters is discretized by utilizing 11 distinct groups of nanoparticles, with the groups' diameters drawn to match the lognormal distribution. After randomly placing the different melanin shell nanoparticles above and below the bulk structure, the shell nanoparticles are brought onto the bulk structure to form a close-packed structure using a conjugate gradient energy minimization technique,[63] implemented in LAMMPS software package.[60] To bring the nanoparticles to the close-packed state, we apply an additional energy potential to all nanoparticles based on the nanoparticle's x, y, and z position such that the nanoparticles further from center of the box (marked as origin) have a higher energy than the nanoparticles near origin. During this step, the nanoparticles interact via the cLJ potential with the Hamaker constant for all pair-wise interactions ($H_{A-A}$, $H_{A-B}$, $H_{B-B}$) initially set to a weak 0.1 $k_BT$ to prevent nanoparticle overlap. In simpler terms, the close-packed structures with shells are generated by biasing the nanoparticles to move towards the origin with the conjugate gradient method used to minimize that energy gradient (creating the close-packed nanoparticle structure). We then maintain that energy gradient while performing a short molecular dynamics (MD) simulation in the NVT ensemble at $T^* = 1.0$ for ~10,000 timesteps with a timestep size of 0.005 $\tau$ ($\tau$ represents simulation time). This short MD simulation allows the nanoparticles to better pack especially at high nanoparticle diameter dispersity. At the end of this process, the bulk structure returned by CREASE has the desired melanin shell layer on the top (+z direction) and bottom (-z direction) to mimic the supraballs with their melanin shell layer.

All computational visualizations were performed using the Visual Molecular Dynamics (VMD) software.[64]

*FDTD Simulations*

The color reflectance of binary mixture supraballs were obtained by performing three-dimensional FDTD calculations using a commercial-grade Ansys Lumerical 2022 R1 FDTD solver (Ansys, Inc.). The reconstructed supraball structures (with nanoparticle positions) were imported into the FDTD toolbox and assigned corresponding material properties (complex refractive index) for silica[65] and melanin[26] chemistries. The real part of the complex refractive index for silica was obtained using the classic Sellmeier dispersion.[65, 66] The imaginary component (extinction coefficient, $\kappa$) of silica's complex refractive index was used from Ref.[57] To mimic the experimental reflectance measurement conditions using a microspectrophotometer that inscribes ~3 μm x 3 μm sensor area to collect reflected light at the center of the supraball, we defined the FDTD simulation region to occupy the same lateral dimensions centered along the reconstructed supraball structures. The simulations were conducted at normal incidence using a broadband plane wave source, propagating along the -Z direction. Boundary conditions were set to periodic in the lateral dimensions (X and Y). The reflectance data was monitored using a Discrete Fourier Transform (DFT) power monitor placed behind the source injection plane. The simulation time (in fs) and boundary condition along the light propagation direction (Z; perfectly matching layer (PML) boundaries) were appropriately chosen such that the electric field decayed before the end of the simulation (auto-shutoff criteria). All the incident light was either reflected, transmitted, or absorbed. Furthermore, a careful stepwise convergence test (on parameters like proximity of PML boundaries, reflection from the PML, mesh sizes and accuracy, and source and monitor placements) was performed to determine the optimal conditions for running numerical simulations.

The following parametric values were used to run optical simulations for the reconstructed structures: a) an auto non-uniform mesh type with a mesh accuracy of 4 (18 mesh points per wavelength) and inner mesh size of ~12 nm for the structural component of the simulation region, b) a broadband plane wave source injection plane set at ~1.75 μm above the surface of the reconstructed structure, c) a stretched-coordinate PML boundary (steep-angle type) with 32 layers in the light propagation direction (Z plane) arranged ~1.25 μm behind the source injection plane, and d) a reflectance DFT monitor placed at ~0.75 μm behind the source injection plane. All the computed reflectance spectra described in this study were obtained by averaging over simulation results of three reconstructed structures, including calculations performed using both *p*- and *s*-polarization states of incident light with error bars representing the standard deviations. These computed reflectance spectra were further analyzed using CIE standards (see ESI **Section 2**).

**Supporting Information**

- SANS contrast matching methodology; color analysis using CIE convention

**Acknowledgements**

C.M.H., A.P., S.S., Z.H., N.C.G., A.J., and A.D. acknowledge financial support from the Air Force Office of Scientific Research (AFOSR) under Multidisciplinary University Research Initiative (MURI) grant (FA 9550-18-1-0142). J-J.S. and S.K.S. appreciate support from the Office of Basic Energy Sciences, U.S. Department of Energy (DOE) under DOE grant No. DE-SC0018086. B.V. and M.D.S. recognize support from AFOSR grant (FA9550-18-1-0477), FWO grant (G007117N), and HFSP grant (RGP 0047). This work was supported with computational resources from the University of Delaware (Farber and Caviness clusters). A.P. and A.D. thank Dr. Mesfin Tsige for providing access to high-performance workstations at the University of Akron. A.P., S.S., J-J.S., M.B., S.K.S., and A.D. gratefully acknowledge the efforts of Cedric Gagnon, Jeff Kryzwon, and John Barker during data collection on the beamlines at NIST. Access to vSANS, NGB 30m SANS, and BT5 uSANS was provided by the Center for High Resolution Neutron Scattering (CHRNS), a partnership between the NIST and the National Science Foundation (NSF) under Agreement No. DMR-1508249. Specific commercial equipment, instruments, or materials are identified in this paper to foster understanding. Such identification does not imply recommendation or endorsement by the NIST, nor does it imply that the materials or equipment identified are necessarily the best available for the purpose. Finally, we extend our gratitude to the entire MURI Melanin team for helpful discussions and insights throughout the course of this project.

**Author Contributions**

[‡]C.M.H. and A.P. contributed equally to this work. C.M.H., A.P., A.J., and A.D. contributed to project conceptualization and design of experiments. C.M.H. conducted CREASE development and structure reconstructions with feedback from A.J., and A.P. performed scattering experiments, optical modeling (FDTD calculations), and data analysis with feedback from A.D. Z.H. and N.C.G. performed synthesis of melanin and silica nanoparticles. A.P. and S.S. fabricated the binary (melanin + silica) mixture supraballs. B.V. and M.D.S. supported the reflectance measurements of the binary mixture supraballs. A.P., S.S., J-J.S. and M.B. performed neutron scattering experiments at the different SANS beamlines and processed/analyzed raw data. S.K.S. provided guidance during neutron scattering experiments. C.M.H. and A.P. wrote the initial manuscript; C.M.H., A.P., A.J., and A.D. reviewed and edited the manuscript. All authors have given approval to the final version of this manuscript.

**Notes**

The authors declare no competing financial interest.